\documentclass[11pt]{article}

\usepackage{graphicx}
\usepackage{epstopdf}
\usepackage{amsmath}
\usepackage{amssymb}
\usepackage{amsfonts}
\usepackage{amsthm}
\usepackage[usenames]{color}
\usepackage{array}
\usepackage{lscape}
\usepackage{xstring}
\usepackage{letltxmacro}

\LetLtxMacro\origcite\cite
\renewcommand{\cite}[1]{%
\begingroup
\def\tempx{0}%
  \StrCount{#1}{,}[\tempx]%
  \ifnum\tempx > 0 
  Refs. %
  \else
  Ref. %
  \fi
\endgroup
\origcite{#1}%
}

\usepackage[
colorlinks=true,
linkcolor=blue,
urlcolor=blue,
filecolor=blue,
citecolor=red,
pdfstartview=FitV,
pdftitle={},
pdfauthor={},
pdfsubject={},
pdfkeywords={},
pdfpagemode=None,
bookmarksopen=true
]{hyperref}

\usepackage{epsfig}

\usepackage{hyperref}
\usepackage{amsmath}

\newcommand{\bea}{\begin{eqnarray}}
\newcommand{\eea}{\end{eqnarray}}
\newcommand{\ba}{\begin{array}}
	\newcommand{\ea}{\end{array}}
\newcommand{\ee}{\end{equation}}

\numberwithin{equation}{section}

\begin{document}

\begin{flushright}
	\texttt{\today}
\end{flushright}

\begin{centering}
	
	\vspace{2cm}
	
	\textbf{\Large{
			  Flat-space Limit   of Extremal Curves }}
	
	\vspace{0.8cm}
	
	{\large   Reza Fareghbal$^{a,b}$, Mehdi Hakami Shalamzari$^{a,c}$, Pedram Karimi$^{a}$ }
	
	\vspace{0.5cm}
	
	\begin{minipage}{.9\textwidth}\small
		\begin{center}
			
			$^a${\it  Department of Physics, 
				Shahid Beheshti University, 
				G.C., Evin, Tehran 19839, Iran.  }\\ 
			$^b${\it  School of Particles and Accelerators, 
                       Institute for Research in Fundamental Sciences (IPM),  
                            P.O. Box 19395-5531, Tehran, Iran }\\ 
            $^c$ {\it Institute for Theoretical Physics, TU Wien
                      Wiedner Hauptstrasse. 8, A-1040 Vienna, Austria}

			\vspace{0.5cm}
			{\tt  r$\_$fareghbal@sbu.ac.ir, fareghbal@theory.ipm.ac.ir }
			\\ \tt $ \, $ma$\_$hakami@sbu.ac.ir \\ \tt  \, pedramkarimie@gmail.com \\				
		\end{center}
	\end{minipage}


	\begin{abstract}
 According to the Ryu-Takayanagi prescription, the entanglement entropy of subsystems in the boundary conformal field theory (CFT) is proportional to the area of extremal surfaces in bulk asymptotically Anti-de Sitter (AdS) spacetimes. The flat-space limit of these surfaces is not well defined in the generic case. We introduce a new curve in the three-dimensional asymptotically AdS spacetimes with a well-defined flat-space limit. We find this curve by using a new vector, which is vanishing on it and is normal to the bulk modular flow of the original interval in the two-dimensional CFT. The flat-space limit of this new vector is well defined and gives rise to the bulk modular flow of the corresponding asymptotically flat spacetime. Moreover, after  Rindler transformation, this new vector is the normal Killing vector of the BTZ inner horizon. We reproduce all known results about the holographic entanglement entropy of Bondi-Metzner-Sachs invariant field theories, which are dual to the asymptotically flat spacetimes. 
	\end{abstract}

\end{centering}

\newpage

\tableofcontents

\section{Introduction}

One of the proposals for the holographic dual of asymptotically flat spacetimes is given by \cite{Bagchi:2010zz,Bagchi:2012cy}. According to this proposal, the flat-space limit in the gravity side in the context of AdS/CFT corresponds to taking the ultrarelativistic limit in the field theory side. For the two dimensional conformal field theories (CFTs), the ultrarelativistic limit of conformal algebra is performed using the Inonu-Wigner contraction and imposing the vanishing speed of light limit. The resultant algebra, in this case, is infinite dimensional, and it is known as Carroll algebra \cite{Leblond65,Duval:2014uoa}. This two-dimensional ultrarelativistic algebra is isomorphic to a three-dimensional relativistic algebra, which is asymptotic symmetry of asymptotically flat spacetimes at null infinity \cite{Barnich:2006av}. The asymptotic symmetry of four-dimensional asymptotically flat spacetimes is known from a long time ago by the works of Bondi, Metzner, and also Sachs, and this symmetry is known as BMS symmetry \cite{BMS}. Recent progress in this way by Barnich \textit{et al}. \cite{Barnich:2006av,Barnich:4dBMS,aspects} shows that by imposing nonglobal invariance of symmetry algebra (which was absent in the first formulation of $bms_4$ algebra in \cite{BMS} and $bms_3$ algebra in \cite{Ashtekar:1996cd}), one can obtain infinite-dimensional algebras in both three- and four-dimensional asymptotically flat spacetimes, which are an extension of translation and rotation of Poincar\'e algebra. In this view, both of $bms_3$ and $bms_4$  consist  of supertranslations and superrotations.

The isomorphism between two-dimensional Carroll algebra and three-dimensional BMS algebra was the motivation of \cite{Bagchi:2010zz,Bagchi:2012cy} to propose that the holographic dual of asymptotically flat spacetimes are indeed BMS-invariant field theories (BMSFT). BMSFTs are ultrarelativistic theories in one-dimensional lower spacetimes characterized by their infinite-dimensional symmetry algebra. In this paper, we call this correspondence flat/BMSFT. The infinite-dimensional symmetry of BMSFTs provides some universal properties which are independent of the detail of these theories. In this view, one can study the flat-space holography by just using the universal properties of BMSFTs  For recent progress in aspects of BMSFTs, see \cite{Bagchi:2019xfx,Bagchi:2019clu}.

There are two approaches to improve the dictionary of flat/BMSFT. One is to study BMSFTs directly and relates its observable to the properties of asymptotically flat spacetimes. Another is to start from AdS/CFT and take a limit from its dictionary. In the bulk side, this limit is simply the flat-space limit or zero cosmological constant limit, which applies by taking the limit of Anti-de Sitter (AdS) radius, $\ell$,  to infinity. 

In this paper, we focus on the second approach to study the holographic entanglement entropy of BMSFTs. because of the infinite-dimensional symmetry of two-dimensional BMSFTs, a universal formula has been proposed for the entanglement entropy in these theories \cite{Bagchi:2014iea}. Later, \cite{Jiang:2017ecm} proposed a holographic description for the BMSFTs entanglement entropy. This holographic description of entanglement entropy is very similar to  Ryu and Takayanagi's (RT) proposal in the context of AdS/CFT \cite{Ryu:2006bv}, which associates the dual field theory entanglement entropy to the length of some extremal curves inside the gravitational theory. In the original RT proposal for the asymptotically AdS spacetime, this curve is a geodesic with extremized length anchored to the end points of the interval at the boundary. In the flat/BMSFT correspondence, this curve is also a geodesic with extremized length, but it connects to two null rays emitted from null infinity where end points of the interval lie. Thus in the flat case, the extremal curves are not connected to the dual field theory intervals directly, and this makes finding them a bit tedious and ambiguous.

The method of \cite{Jiang:2017ecm} for constructing extremal curves in asymptotically flat spacetimes is based on  Rindler transformation \cite{Faulkner:2013ica}. In the field theory side, this unitary transformation is part of the symmetry of the theory, which relates the original entanglement entropy to a thermal entropy. The thermal entropy corresponds to the entropy (area) of an event horizon in the gravitational theory. The spacetime, in which this event horizon exists, is given by the bulk Rindler transformation from original spacetime. The extremal curve in the original bulk spacetime is nothing but the inverse Rindler transformation of the final event horizon. Thus in this method, one needs to find  Rindler transformation and its bulk extension for finding extremal curves. In this view, we do not need to extremize any length, and the extremal curve is a result of  Rindler transformation. Although this method applies to all field theories, finding corresponding Rindler transformation is not straightforward.

In the original RT  proposal in the context of AdS/CFT, it is not necessary to know Rindler transformation. Imposing extremality conditions for the length of bulk geodesics anchored to the end points of the boundary interval is enough to determine the RT curve. One can interpret the flat-space limit of these curves as the holographic entanglement entropy related to BMSFTs. However, the flat-space limit of these curves is not well defined. The simplicity of the extremization procedure, which is used in the AdS/CFT case, is tedious enough to make someone think about the extremal curves with a well-defined flat-space limit. In this paper, we focus on this problem and introduce a new curve in the asymptotically AdS spacetimes whose flat-space limit yields the results of \cite{Jiang:2017ecm}.

The main idea in the current calculation is related to the observation of \cite{Fareghbal:2014qga} and \cite{Riegler:2014bia}  that the entropy of flat-space cosmological solution (FSC) which is given by taking the flat-space limit from the BTZ black hole \cite{Cornalba:2002fi} \cite{Barnich:2012xq}, is a result of flat-space limit performed on the entropy of BTZ inner horizon. After  Rindler transformation, the RT extremal curve transforms into the outer horizon of the corresponding black hole, whereas the curve with a well-defined flat-space limit transforms to the inner horizon.

To bypass  Rindler transformation and introduce a method that solely uses the extremality condition, we use the bulk modular flow of the corresponding interval in the AdS case. It vanishes on the RT extremal curve and transforms into the Killing vector normal to the outer horizon after  Rindler transformation. The fact that our new curve relates to the inner horizon reveals that the Killing vector normal to the inner horizon can be meaningful in the original coordinate if we apply the inverse Rindler transformation. The interesting point is that the Killing vectors normal to inner and outer horizons are perpendicular to each other, and the sum of their norms is constant. Thus, besides the Killing equation, we have two other equations that associate two vector fields to each other in any coordinate system. One of these vector fields is the bulk modular flow of the original coordinate in the AdS case, and the second one is a new vector field that has a well-defined flat-space limit. The flat-space limit of our new vector field in the asymptotically AdS spacetime yields the bulk modular flow of the corresponding interval introduced in \cite{Jiang:2017ecm} for BMSFT. The interesting point is that the components of the new vector field vanish on the new extremal curve. In other words, this new curve consists of the fixed points of the new vector. Thus, we can bypass  Rindler transformation in this way by starting from the bulk modular flow of the original interval in the AdS case and constructing a new vector field by using the Killing equations, normality condition, and norm condition. These equations determine our new vector field. Then, we can take the flat-space limit and find the bulk modular flow of the asymptotically flat case and finally obtain the extremal curve introduced in \cite{Jiang:2017ecm}.

In the holographic description proposed for the entanglement entropy of BMSFT introduced in \cite{Jiang:2017ecm}, besides an extremal spacelike curve, two null curves are starting from null infinity from end points of the  BMSFT interval and intersect the extremal curve. This is the length of the extremal curve bounded between two intersections by null curves that is proportional to the BMSFT entanglement entropy. In this paper, we also propose two new null curves in the asymptotically AdS spacetimes, which their flat-space limit results in the null rays of \cite{Jiang:2017ecm}. The fact that our new curve gets to the boundary at points which are different than the end points of the required interval makes it possible to find null rays that connect end points of the interval to the new curve. However, the number of these null geodesics is infinite. We observe that imposing the condition that null curves also pass through two cutoff points picks up two distinct null curves that their flat space limit yields the flat null rays of \cite{Jiang:2017ecm}.

In the original paper by Casini \textit{et al.} \cite{Faulkner:2013ica}, it is shown that Rindler transformation is such that it maps the causal development of a spherical entangling region on a CFT to a new thermal CFT on a hyperbolic space (cross a circle in complex time). Then, by the AdS/CFT duality, this new thermal CFT is dual to the exterior of a topological black hole. Thus, geometrically, the map can be taken, at the bulk level, as a map between the region contained within the RT surface of a sphere in pure AdS and the exterior of a topological black hole in AdS space. Therefore, it is a map between the interior of the RT surface and the exterior of a black hole, such that the RT surface itself maps to the outer black hole horizon. In this picture, the region bounded by the interior of the RT surface and the exterior region of the black hole are AdS
spacetimes that extend to the conformal boundary.  In the setup of this paper, the new RT-like extremal curve gets mapped under Rindler transformation to the inner horizon of the BTZ black hole instead of the outer horizon. Since the inner horizon is not in causal contact with the exterior of the black hole, it is unclear in the first view to what region in the black hole the bounded region by this new RT curve will be mapped to. This ambiguous point can be resolved by the fact that after the flat-space limit the outer horizon of the BTZ black hole gets a map to infinity, and the region between the outer horizon and conformal boundary vanishes after taking this limit. Thus, the region contained within the new RT-like surface and the AdS boundary maps to the region between the inner and outer horizons. Although this region is causally disconnected from the conformal boundary in the AdS case, after taking the flat-space limit, the extra region between the outer horizon and the conformal boundary disappears, and consequently, both areas before and after Rindler transformation are casually connected to null infinity.

In Sec. II, we start from preliminaries and review relevant topics necessary for the rest of this paper. Section III is the central part of our paper and consists of the calculations for introducing the new curves and new vector field with a well-defined flat-space limit. We also propose an altered recipe for the holographical calculation of BMSFT entanglement entropy, which does not require Rindler transformation.  The last section is devoted to the summary and conclusion.

\section{Preliminaries}

\subsection{Entanglement entropy, modular Hamiltonian, modular flow and Rindler transformation}
For the subsystem, $\mathbf{A} $, of a quantum field theory in a pure state $|\Psi \rangle$, the reduced density matrix is given by
\begin{equation}
	\rho_{\mathbf{A}} = \text{Tr}_{\bar{\mathbf{A}}} ~\rho = \text{Tr}_{\bar{\mathbf{A}}} ~ |\Psi \rangle \langle \Psi |,
\end{equation}
where  $\bar{\mathbf{A}}$ is the complement of $\mathbf{A}$.
The entanglement entropy of region $\mathbf{A}$ is given by the von-Neumann entropy of the  reduced density matrix
\begin{equation}\label{entropyqm}
	S_{\mathbf{A}} = - \text{Tr} ( \rho_{\mathbf{A}} ~ \log{\rho_\mathbf{A}}). 
\end{equation}
 Since $\rho_{\mathbf{A}}$ is Hermitian and positive semidefinite, we can write it in term of another Hermitian operator $\mathcal{H}_\mathbf{A}$:
\begin{equation}
	\rho_{\mathbf{A}} = \frac{e^{-\mathcal{H}_\mathbf{A}}}{ \text{Tr} ( e^{- \mathcal{H}_{\mathbf{A}} } )}.
\end{equation}
$\mathcal{H}_\mathbf{A}$ is called the entanglement Hamiltonian or more commonly modular Hamiltonian in the literature and is the conserved charge of a geometric flow $k_t$, which is known as modular flow.

For a generic QFT, the calculation of the entanglement entropy given by \eqref{entropyqm} is not straightforward. However, this calculation is simplified by making use of the symmetry of the QFT.

 There are two methods for this calculation, which are known as the replica trick \cite{Holzhey:1994we,Calabrese:2004eu} and the  Rindler method \cite{Faulkner:2013ica}. Here, we briefly introduce the Rindler method since it is generalized to the BMSFTs. 
 
 The Rindler method aims to find symmetry transformation, $U_R$, that maps the density matrix of the entangled region to the thermal one,
\begin{equation} \label{eetoth}
	\rho_{A} = U_R ~\rho_{\tilde{B}}~ U^{-1}_{R},
\end{equation}
where the tilde entities stand for the thermal system. Since the unitary transformation does not change the entropy of the system, we would expect that entanglement entropy of the region $\mathbf{A}$ to be equal to the thermal entropy of the region $\tilde{B}$. 

  Now in the thermal system $\tilde{B}$, we can identify the partition function and the geometric flow
 \begin{equation}
	 Z(\tilde{B}) = \text{Tr} ~e^{- \tilde{\beta}^i Q_{\tilde{x}_i }}, \qquad \rho_{\tilde{B}} =Z(\tilde{B})^{-1} e^{- \tilde{\beta}^i 						Q_{\tilde{x}^i}}.
 \end{equation}
Identifying the partition function for a thermal system, we can work out the entropy as well,
 \begin{equation}
 S(A)_{EE} = S(\tilde{B}) = (1-\tilde{\beta}^i \partial_{\tilde{\beta}^i} ) Z(\tilde{B}).
 \end{equation}
  The  modular flow can be written as  $k_{t} = \tilde{\beta}^{i} \partial_{\tilde{x}^i}$. It vanishes at the boundary of the entangled surface, $k_t |_{\partial{\mathbf{A}}} = 0$ \cite{Jiang:2017ecm}.
 
 \subsection{Holographic entanglement entropy in AdS/CFT and bulk modular flow} 
In the gauge/gravity duality, the entanglement entropy of the boundary subsystems has a geometric interpretation in terms of extremal curves within the bulk theory introduced by RT in the context of AdS/CFT \cite{Ryu:2006bv}. Accordingly, the holographic entanglement entropy is given by 
\begin{equation}\label{RT}
S = \frac{\mathcal{A}}{4 G_N},
\end{equation}
where $\mathcal{A}$   is an extremal surface anchored to the interval ${\mathbf{A}}$ on the conformal boundary of asymptotically local AdS spacetime and $G_N$ is Newton's constant. 

Let us consider AdS$_3$, with  line element
\begin{equation}\label{fgmetric}
	{ds}^2 = \frac{\ell^2}{z^2} ( dx^2+dz^2-dt^2),
\end{equation}
where its conformal boundary is given by $z=0$. For an  interval at the boundary of this spacetime addressed by 
\begin{align}\label{interval poin}
{\mathbf{A}}: \begin{cases}
							&-R<x<R \\
							&t =0,
						 \end{cases}
\end{align}
 the extremal surface in bulk is given by
\begin{equation} \label{RTFG}
x^2+z^2 =R^2,\qquad t=~0
\end{equation}  
 So using \eqref{RT}, we find that 
\begin{equation}
	S = \frac{2\ell}{4   G_N} \int_{\epsilon}^{R} {\frac{dz}{z  \sqrt{ R^2 - z^2 }} }= \frac{\ell}{2   G_N} \log{ \frac{2   R}{\epsilon}};
\end{equation}
here $\epsilon $ is a cutoff. This is the celebrated entanglement entropy in $CFT_2$ \cite{Holzhey:1994we,Calabrese:2004eu}.

We can also find the bulk modular flow $k_t^{bulk}$ using the extension of the boundary modular flow to bulk.  The modular flow of the interval \eqref{interval poin} is given by \cite{Faulkner:2013ica}
\begin{equation}\label{boundary modular flow poi}
	k_t=-\dfrac{2\pi x t}{R}\partial_x+\dfrac{\pi}{R}\left(R^2-t^2-x^2\right)\partial_t.
\end{equation}
 The RT extremal surface, $\mathcal{A}$,  can be interpreted as the surface at which
 \begin{equation}\label{Condition of bulk modular flow}
	 k_t^{bulk}\big|_\mathcal{A} = 0.
 \end{equation}
 Moreover, $k_t^{bulk}$ is a Killing vector of the bulk metric and $k_t^{bulk}\big|_{z=0}=k_t$. Putting all together, we find
\begin{equation}\label{bulk modular poin ads}
	k_t^{bulk}\equiv\xi=- \dfrac{2\pi  t}{R}\left( x\partial_{x}+z\partial_z\right) +\dfrac{\pi}{R}\left( R^2 -x^2 -t^2 - z^2\right ) \partial_{t}.
\end{equation}

Since the general interval plays a crucial role in the rest of this paper, we also calculate the modular flow of the boosted interval. To this aim, we boost the boundary coordinates 
\begin{align}\label{boost}
						\nonumber	t^\prime &= t  \cosh{\eta} + x \sinh{\eta},\\
							x^\prime &= t \sinh{\eta} + x \cosh{\eta}.
\end{align}
Demanding the new interval to satisfy
\begin{align}
{\mathbf{\Delta}}': \begin{cases}
							&\frac{-l_x}{2}<x'<  \frac{l_x}{2},\\
						 	&\frac{-l_t}{2}<t' < \frac{l_t}{2},\\
							&R^2 =\frac{l_x^2-l_t^2}{4} ,~(l_x > l_t),
						 \end{cases}
\end{align}
will uniquely determine the boost parameter $\eta$ as
\begin{equation}
	\eta=\log\sqrt{\dfrac{l_x^2+l_t^2}{l_x^2-l_t^2}}.
\end{equation}

 The modular flow for this boosted interval can be obtained using \eqref{boost} 
\begin{align}\label{boosted modular flow} 
	\nonumber\xi=&  \frac{\pi \left(l_x^2  l_t  - l_t^3  -8 l_x x' t' + 4 l_t (x'^2 + t'^2  - z^2)\right)}{2(l_x^2  - l_t^2) } \partial_{x'} 
	\\
	&+  \frac{\pi \left(l_x^3 + 8 l_t x' t'  - l_x l_t^2  - 4 l_x (x'^2 + t'^2 + z^2)\right)}{2 (l_x^2  - l_t^2) } \partial_{t'} -\frac{4 \pi (- l_t x' + l_x t') z}{l_x^2  - l_t^2 } \partial_z.
\end{align}
For the boosted interval, the holographic entanglement entropy is given by the following extremal surface
\begin{align}
			 x'^2 =  \frac{l_x^2(l_x^2 - l_t^2 - 4 z^2)}{4(l_x^2-l_t^2)},\qquad
			y'^2 =  \frac{l_t^2(l_x^2 - l_t^2 - 4 z^2)}{4(l_x^2-l_t^2)}.
\end{align}

\subsection{Brief review of flat/BMSFT correspondence }
Asymptotically flat spacetimes are given by taking the flat-space limit (zero cosmological constant limit or large AdS radius limit) from the asymptotically AdS spacetimes. According to the proposal of \cite{Bagchi:2012cy}, the flat-space limit in the asymptotically local AdS spacetime corresponds to the ultrarelativistic limit in the boundary CFT. The resultant ultrarelativistic field theory is known as BMSFT, and the correspondence between asymptotically flat spacetimes and BMSFTs is called flat/BMSFT. BMSFTs are BMS-invariant field theories. For the $d$-dimensional BMSFTs, BMS symmetry is the asymptotic symmetry of $(d+1)$-dimensional asymptotically flat spacetimes at null  nfinity. The BMS algebra as the asymptotic symmetry algebra is infinite dimensional. In three dimensions, bms$_3$ is given by 
\begin{equation}\label{bmsa}
	\begin{split}
		&[L_m , L_n] =(m-n) L_{m+n}, \\
		&[L_m , M_n] =(m-n) M_{m+n},
		\\
		&[M_m , M_n] = 0, \qquad\qquad\qquad m,n \in \mathbb{Z}.
	\end{split}
\end{equation} 
where $L_n$ and $M_n$ are, respectively, the generators of superrotation and supertranslation. For $n=0,-1,1$ the resultant subalgebra is Poincar\'e algebra. The algebra \eqref{bmsa} is the asymptotic symmetry algebra of three-dimensional spacetimes given by  
\begin{equation}\label{afm}
	ds^2 = {M}du^2 -2du dr +2 {N}du d\phi +r^2 d\phi^2,
\end{equation}
where ${M}$ and ${N}$ are functions of $u$ and $\phi$ and they satisfy 
\begin{equation}
	\partial_u {M} = 0 \ , \qquad 2 \partial_u {N} = \partial_{\phi} {M}. 
\end{equation}
$u$ is retarded time and coordinate ${u,r,\phi}$  in \eqref{afm} is known as the BMS coordinate \cite{Barnich:2012aw}. The line element \eqref{afm} is given by taking the flat-space limit from the  asymptotically AdS metric, 
\begin{equation}\label{BMS gauge}
	ds^2 = \left( - \frac{r^2}{\ell^2} + \mathcal{M}\right)du^2 -2du dr +2 \mathcal{N}du d\phi +r^2 d\phi^2,
\end{equation} 
where  $\mathcal{M}$ and $\mathcal{N}$ are functions of $u$ and $\phi$ and are constrained  by using the equations of motion as 
\begin{equation}
	\partial_u \mathcal{M} = \frac{2}{\ell^2} \partial_{\phi}\mathcal{N} \ , \qquad 2 \partial_u \mathcal{N} = \partial_{\phi} \mathcal{M}. 
\end{equation}

The functions $M$ and $N$ are the resultant functions of taking the flat-space limit from the functions $\mathcal{M}$ and $\mathcal{N}$.

The generators of \eqref{bmsa} can be obtained by taking the flat-space limit from the generators of conformal algebra \cite{Barnich:2012aw},
\begin{equation}\label{flat limit of generators}
	L_m = \lim\limits_{\frac{G}{\ell}\rightarrow 0} \, (\mathcal{L}_m-\bar{\mathcal{L}}_{-m}), \qquad M_m =  \lim\limits_{\frac{G}{\ell}\rightarrow 0} \,\frac{G}{\ell} (\mathcal{L}_m+\bar{\mathcal{L}}_{-m}).
\end{equation}
where $\mathcal{L}_m$ and $\bar{\mathcal{L}}_m$ are the generators of the conformal algebra, 
\begin{equation}\label{CA}
	\begin{split}
		&[\mathcal{L}_m , \mathcal{L}_n] =(m-n) \mathcal{L}_{m+n}, \\
		&[\bar{\mathcal{L}}_m , \bar{\mathcal{L}}_n] =(m-n) \bar{\mathcal{L}}_{m+n},
		\\
		&[\mathcal{L}_m , \bar{\mathcal{L}}_n] = 0, \qquad\qquad\qquad\qquad m,n \in \mathbb{Z}.
	\end{split}
\end{equation}
It was proposed in \cite{Bagchi:2012cy} that the limit \eqref{flat limit of generators}, which is taken in the gravity side, corresponds to the ultrarelativistic limit in the field theory side. The algebra of conserved charges in both of \eqref{bmsa} and \eqref{CA} are centrally extended. 

Similar to other field theories, it is possible to define the entanglement entropy for the subsystems of BMSFT. The infinite-dimensional symmetry of BMSFTs admits to finding universal formulas for the entanglement entropy of subregions \cite{Bagchi:2014iea}. Moreover, using the flat/BMSFT correspondence, one can find a holographic description for the BMSFT entanglement entropy. Recently, a prescription (similar to the   Ryu-Takayanagi proposal for the CFT entanglement entropy \cite{Ryu:2006bv})  has been proposed for the BMSFT entanglement entropy  \cite{Jiang:2017ecm} that relates it to the area of some particular curves in the asymptotically flat bulk spacetimes. (See also \cite{Wen:2018mev} \cite{Wen:2018whg}.)

To be precise, let us consider the null-orbifold or Poincar\'e patch [with ${M}= {N}=0$ in \eqref{afm}], which is given by
\begin{equation}\label{poincare flat}
	ds^2=-2dudr+r^2 d\phi^2.  
\end{equation}
For the interval $B$ in the dual BMSFT that identifies by $-\frac{l_u}{2}<u<\frac{l_u}{2}$ and $-\frac{l_{\phi}}{2}<\phi<\frac{l_{\phi}}{2}$ where $l_u$ and $l_{\phi}$ are constants, the entanglement entropy is \cite{Jiang:2017ecm}
\begin{equation}\label{EE Poincare flat}
	S_{EE}={C_{LL}\over6}\log\dfrac{l_\phi}{\epsilon_\phi}+{C_{LM}\over 6}\left(\dfrac{l_u}{l_\phi}-\dfrac{\epsilon_u}{\epsilon_\phi}\right)
\end{equation}
where $\epsilon_\phi$ and $\epsilon_u$ are cutoffs in $\phi$ and $u$ directions and $C_{LL}$ and $C_{LM}$ are central charges of \eqref{bmsa} related to the central charges $c$ and $\bar c$  of conformal algebra \eqref{CA} as
\begin{equation}\label{flat limit of cc}
	C_{LL} = \lim\limits_{\frac{G}{\ell}\rightarrow 0} \, (c-\bar c), \qquad C_{LM} =  \lim\limits_{\frac{G}{\ell}\rightarrow 0} \,\frac{G}{\ell} (c+\bar c).
\end{equation}
For the BMSFT dual to Einstein gravity, $C_{LL}=0$ and $C_{LM}=3$. 

 According to \cite{Jiang:2017ecm}, the  entanglement entropy of subregion $B$ of BMSFT$_2$ is given by
\begin{equation}\label{song formula}
	S_{HEE} = \frac{\text{Length}(\gamma)}{4G} = \frac{\text{Length}(\gamma \cup \gamma_+ \cup \gamma_-)}{4G}
\end{equation}
where $\gamma$ is a spacelike geodesic and $\gamma_+$ and $\gamma_-$ are null rays from $\partial\gamma$ to $\partial B$. To find these curves, \cite{Jiang:2017ecm} uses a Rindler transformation as
\begin{eqnarray}\label{Rindler poin flat}
 	\nonumber\tilde r &=& \sqrt{{{\tilde M}\over 16 l_\phi^2}\Big(8u-4l_u+r(l_\phi-2\phi )^2\Big) \Big(8u+4l_u+r(l_\phi+2\phi )^2\Big) +{{\tilde J}^2\over 4 \tilde M}} \,,\\
	\nonumber\tilde \phi &=& -\frac{1}{\sqrt{\tilde{M}}} \log {\sqrt{\tilde M}\over 4 l_\phi} \Big(\frac{8  u-4l_u+r(l_\phi-2\phi )^2} {{\tilde r}+{\tilde J /( 2\sqrt{\tilde M})} }\Big)\,,
	 \\
	\tilde u &=& {\tilde r \over \tilde M}+ {1\over 4l_\phi\sqrt{\tilde M}}\big( 8u+4 r\phi^2-r l_\phi^2\big)  -{\tilde J\over 2\tilde M} \tilde \phi\,.
\end{eqnarray} 
which transforms \eqref{poincare flat} to 
\begin{equation}\label{FSC after Rindler}
	ds^2= \tilde M d\tilde u^2-2 d\tilde ud\tilde r +\tilde J d\tilde ud\tilde \phi+\tilde r^2 d\tilde\phi^2\
\end{equation}
where $\tilde M$ and $\tilde J$ are constants. The metric \eqref{FSC after Rindler} is known as FSC \cite{Cornalba:2002fi, Cornalba:2003kd}  and has a cosmological horizon located at 
\begin{equation}
	\tilde r_C=\dfrac{\tilde J}{2\sqrt{\tilde M}}.
\end{equation}
If we assume that $\gamma$ and $\gamma_\pm$ are mapped to the cosmological horizon after the bulk Rindler transformation \eqref{FSC after Rindler}, one can start with the condition $\tilde r=\tilde r_C$ and use \eqref{Rindler poin flat} to find them \cite{Jiang:2017ecm}. In the next section, we find these curves by making a flat-space limit from particular curves in the asymptotically AdS spacetimes.

\section{Holographic BMSFT entanglement entropy using flat-space limit}
To study the holographic entanglement entropy of BMSFTs by using the method of \cite{Jiang:2017ecm}, one needs to find appropriate Rindler transformation. On the other hand, taking the flat-space limit from the known results in the AdS/CFT correspondence is another method for studying flat-space holography. In this section, we propose proper curves in the asymptotically AdS spacetimes, in which their flat-space limit results in $\gamma$ and $\gamma_\pm$. We restrict our study to the AdS$_3$ in the Poincar\'e coordinate, and the Appendix argues that our results are easily generalizable to other coordinates by making use of coordinate transformation. In other words, we propose an alternative method for studying the holographic entanglement entropy of BMSFTs, which does not use Rindler transformation
\subsection{Initial setup in three-dimensional asymptotically AdS spacetime}

The main goal is to take the flat-space limit from the holographic calculation in bulk, which is  AdS$_3$ spacetime in Poincar\'e coordinate. However, to have a well-defined flat-space limit, we need to write our metric in an appropriate gauge. By well-defined gauge, we mean a set of the coordinate system where taking the flat-space limit $\ell \to \infty$ of the $AdS_3$ metric ends up with well-defined three-dimensional (3d) flat-space metric. This appropriate  gauge was introduced in \cite{Barnich:2012aw} and is called the BMS gauge. 

The  $AdS_3$ in Poincar\'e-BMS coordinate  is  written as
\begin{align}\label{poincare-bms}
	ds^2= r^2 \left(d\phi ^2-\frac{{du}^2}{\ell ^2}\right)-2 {dr} {du}.
\end{align}
This metric is given by the following transformation from the metric \eqref{fgmetric},
\begin{align}\label{poin to bms ads}
	\nonumber&x = \ell\phi,\\
	\nonumber&t=u-\dfrac{\ell^2}{r},\\
	&z=\dfrac{\ell^2}{r}.
\end{align}
This coordinate transformation gives rise to a well-defined flat limit of the interval as well as the metric. Using \eqref{boosted modular flow} and \eqref{poin to bms ads}, we can find the components of the  bulk modular flow,
\begin{align}\label{bulk modular flow BMS ads}
	\nonumber& \xi^r =-\frac{4 \pi  \ell  \left(l_{\phi } \left(\ell ^2-r u\right)+r \phi  l_u\right)}{l_u^2-\ell ^2 l_{\phi }^2},\\
	\nonumber& \xi^u =\frac{\pi  \ell  \left(-l_{\phi } \left(l_u^2+4 \left(u^2+\ell ^2 \phi ^2\right)\right)+8 u \phi  l_u+\ell ^2 l_{\phi }^3\right)}{2 \left(l_u^2-\ell ^2 l_{\phi }^2\right)},\\
	& \xi^\phi =\frac{\pi  \left(l_u \left(r \ell ^2 l_{\phi }^2+4 r \left(u^2+\ell ^2 \phi ^2\right)-8 u \ell ^2\right)+8 \ell ^2 \phi  l_{\phi } \left(\ell ^2-r u\right)-r l_u^3\right)}{2 r \ell  \left(l_u^2-\ell ^2 l_{\phi }^2\right)},
\end{align}
where $l_u$ and $l_\phi$ are 
\begin{align}
 l_u=l_t ,\qquad  l_\phi=\dfrac{l_x}{\ell} .
\end{align}
Taking  the $r\to\infty$ limit from $\xi^u$ and $\xi^\phi$ in  \eqref{bulk modular flow BMS ads} results in the components of the modular flow for an interval in the boundary CFT, which is determined by $-\frac{l_u}{2}<u<\frac{l_u}{2}$ and $-\frac{l_\phi}{2}<\phi<\frac{l_\phi}{2}$.

\subsection{New vector normal to bulk modular flow }

Although the flat-space limit of \eqref{poincare-bms} is well defined and gives rise to \eqref{poincare flat}, it is not difficult to check that the $\ell\to\infty$ limit is not well defined for \eqref{bulk modular flow BMS ads}. This means that we cannot find the modular flow of a similar interval in BMSFT by taking the flat-space limit from the modular flow of a corresponding interval in CFT. However, starting from \eqref{bulk modular flow BMS ads}, we introduce a new vector field, $\lambda^\mu$, with a well-defined flat-space limit. To do so, we first use Rindler transformation; however, in the end, we introduce a recipe for the calculation of $\lambda^\mu$ from $\xi^\mu$ directly.

The bulk Rindler transformation, which changes \eqref{poincare-bms} to the BTZ black hole written in the BMS gauge, is given in the Appendix. The final metric is 
\begin{equation}\label{ads metric after rindler}
	ds^2 = \left( - \frac{\hat r^2}{\ell^2} + \mathcal{\hat M}\right)d\hat u^2 -2d\hat u d\hat r +2 \mathcal{\hat N}d\hat u d\hat\phi +\hat r^2 d\hat\phi^2,
\end{equation}
where $\mathcal{\hat M}$ and  $\mathcal{\hat N}$ are given in terms of the BTZ inner and outer horizons $\hat r_\pm$ as
\begin{equation}
	\mathcal{\hat M}=\dfrac{\hat r_+^2+\hat r_-^2}{\ell^2},\qquad \mathcal{\hat N}=\dfrac{\hat r_+\hat r_-}{\ell}.
\end{equation}
After  Rindler transformation , the components of $\xi^\mu$ in \eqref{bulk modular flow BMS ads} transform as 
\begin{equation}\label{bulk modular flow after rindler in bms}
	\hat\xi^{\hat{\phi}} = \pi \ell \left(\frac{2\hat r_{-}}{\hat r_{+}^2- \hat r_{-}^2}\right), \qquad \hat\xi^{\hat{r}} = 0, \qquad \hat \xi^{\hat{u}} = - \pi \ell^2 \left(\frac{2 \hat r_{+}}{\hat r_{+}^2- \hat r_{-}^2}\right).
\end{equation}
The interesting point is that $\hat \xi^\mu$ is the Killing vector normal to the outer horizon and can be written as
\begin{equation}\label{hat xi }
	\hat \xi= \dfrac{2\pi}{\hat\kappa_+}\left(\partial_{\hat u}-\hat\Omega_+\partial_{\hat\phi}\right),
\end{equation}
where $\hat\kappa_+$ and $\hat\Omega_+$, which are, respectively, the surface gravity and angular velocity of the outer horizon, are given as
\begin{equation}
	 \hat\kappa_+=\dfrac{|\hat r_+^2-\hat r_-^2|}{\ell^2\hat r_+},\qquad \hat\Omega_+=\dfrac{\hat r_-}{\ell \hat r_+}.
\end{equation} 
In all of the previous works, which develop the flat/BMSFT correspondence by taking the flat-space limit from the AdS/CFT calculations, the corresponding quantity in the asymptotically AdS spacetime with a well-defined flat-space limit is related to the inner horizon of the BTZ black hole \cite{Fareghbal:2014qga, Riegler:2014bia}. We want to continue this idea for the current problem and introduce $\lambda^\mu$ as the vector field, which is given by taking the inverse Rindler transformation from the Killing vector field normal to the inner horizon. This vector field which is denoted by $\hat\lambda^\mu$, is given by

\begin{equation}\label{hat lambda }
	\hat \lambda= \dfrac{2\pi}{\hat\kappa_-}\left(\partial_{\hat u}-\hat\Omega_-\partial_{\hat\phi}\right),
\end{equation}
where $\hat\kappa_-$ and $\hat\Omega_-$ are  the surface gravity and angular velocity of the inner horizon,
\begin{equation}
	 \hat\kappa_-=\dfrac{|\hat r_+^2-\hat r_-^2|}{\ell^2\hat r_-},\qquad \hat\Omega_-=\dfrac{\hat r_+}{\ell\hat r_-}.
\end{equation}
Comparing  \eqref{hat xi } and \eqref{hat lambda } shows that they are deducible from each other if we use the following transformation:
\begin{equation}\label{rp to rm}
	\hat r_+\Longleftrightarrow \hat r_-.
\end{equation}

Using the inverse Rindler transformation given in the Appendix, we find $\lambda^\mu$ as 
 \begin{align}\label{lambda before limit poin bms}
	\nonumber & \lambda^r =\frac{4 \pi  \left(l_u \left(\ell ^2-r u\right)+r \ell ^2 \phi  l_{\phi }\right)}{l_u^2-\ell ^2 l_{\phi }^2},\\
	\nonumber & \lambda^u =\frac{\pi  \left(l_u \left(\ell ^2 l_{\phi }^2+4 \left(u^2+\ell ^2 \phi ^2\right)\right)-8 u \ell ^2 \phi  l_{\phi }-l_u^3\right)}{2 \left(l_u^2-\ell ^2 l_{\phi }^2\right)},\\
	& \lambda^\phi =\frac{\pi  \left(-l_{\phi } \left(r l_u^2+4 r \left(u^2+\ell ^2 \phi ^2\right)-8 u \ell ^2\right)+8 \phi  l_u \left(r u-\ell ^2\right)+r \ell ^2 l_{\phi }^3\right)}{2 r \left(l_u^2-\ell ^2 l_{\phi }^2\right)}.
\end{align}
Now, $\ell\to\infty$ is well defined and results in the bulk modular flow  corresponding to the interval in the BMSFT introduced in \cite{Jiang:2017ecm}:
\begin{equation}
	\nonumber\xi_{flat}^{u} = \frac{-\pi (-8 l_{\phi} u \phi+l_u (4\phi^2+l_{\phi}^2))}{2 l_{\phi}^2}, \qquad \xi_{flat}^{r} = \frac{-4 \pi (l_u + l_{\phi} r \phi)}{l_{\phi}^2}, 
\end{equation}
\begin{equation}\label{modular flow flat poincare}
	\xi_{flat}^{\phi} = \frac{-\pi(r l_{\phi}^3-8 l_u \phi+8 u l_{\phi}-4 r l_{\phi}\phi^2)}{2 r l_{\phi}^2}.
\end{equation}

All of $\xi^\mu$, $\lambda^\mu$, $\hat \xi^\mu$, and $\hat \lambda^ \mu$ are Killing vectors of the related spacetimes. Since $\hat \xi\mu$ and $\hat \lambda \mu$ are Killing vectors normal to the horizons, we find that
\begin{align}
	\nonumber\hat\xi^\mu\,\hat\lambda_\mu&=0,\\
	\hat\xi^\mu\,\hat\xi_\mu+\hat\lambda^\mu\,\hat\lambda_\mu&=4\pi^2\ell^2.
\end{align}
Using the fact that $\xi^\mu$ and $\lambda^\mu$ are given by coordinate transformation from $\hat \xi^\mu$ and $\hat \lambda^ \mu$, we can conclude that
\begin{align}\label{equation of lambda}
	\nonumber\xi^\mu\,\lambda_\mu&=0,\\
	\xi^\mu\,\xi_\mu+\lambda^\mu\,\lambda_\mu&=4\pi^2\ell^2.
\end{align}
For a given $\xi$, Eqs. \eqref{equation of lambda} and the fact that $\lambda$ is a Killing vector are enough to determine it without using  Rindler transformation.

 Comparing \eqref{bulk modular flow BMS ads} and \eqref{lambda before limit poin bms}, one can find an interesting relation between the components of $\xi^\mu$ and $\lambda^\mu$. The similar components are changed to each other if we make the following transformation:
 \begin{align}\label{transformation of l}
	  l_u\Rightarrow \ell\,l_\phi,\qquad  l_\phi\Rightarrow\dfrac{l_u}{\ell}.
\end{align}
One can use this simple transformation to find $\lambda^\mu$ of more complicated cases from $\xi^\mu$. In this view, the bulk modular flow of asymptotically flat spacetimes simply achieves from the bulk modular flow of corresponding asymptotically AdS spacetimes by first using the transformation \eqref{transformation of l} and then taking the flat-space limit. Using  Rindler transformation introduced in the Appendix, one can  deduce  \eqref{transformation of l}  from \eqref{rp to rm}.

\subsection{New extremal curve with well-defined flat-space limit}

In this subsection, we execute the final step and find the corresponding curves in the asymptotically AdS spacetimes, of which their flat-space limit results in extremal curves $\gamma$ and $\gamma_\pm$ in the asymptotically flat spacetimes. To this end, let us first look at $\xi^\mu$ and extremal curves in the asymptotically AdS spacetimes whose length, according to the proposal of RT, gives rise to the entanglement entropy of the corresponding interval in the boundary CFT.

To be precise, let us consider an interval in CFT dual to \eqref{poincare-bms}, characterized by $-\frac{l_u}{2}<u<\frac{l_u}{2}$ and $-\frac{l_\phi}{2}<\phi<\frac{l_\phi}{2}$. According to the proposal of \cite{Ryu:2006bv}, the entanglement entropy of this interval is proportional to the length of an extremal curve in the bulk anchored to the end points of the interval located at $(u=-l_u/2,\phi=-l_\phi/2)$ and $(u=l_u/2,\phi=l_\phi/2)$. Thus, using the metric of the bulk, one can find this extremal curve. Moreover, the components of the bulk modular flow $\xi^\mu$ are vanishing on this curve, and it consists of the fixed points of the bulk modular flow. Using \eqref{bulk modular flow BMS ads}, we observe that this extreme curve satisfies the following equation\footnote{Since finally, we  want to  take the  $\ell\to\infty$ limit, it is assumed that $\ell^2l_\phi^2>l_u^2 $}:
\begin{eqnarray}\label{equation for xi}
	\nonumber &\phi l_u-u l_\phi+\dfrac{\ell^2 l_\phi}{r}=0,\\
	& l_\phi l_u^2+4 u^2  l_\phi+ 4\ell ^2 l_\phi \phi ^2-8 u \phi  l_u-\ell ^2 l_{\phi }^3=0.
\end{eqnarray}

 One can also  look for the curve in bulk on which the components of  $\lambda^\mu$ are vanishing. For   $\lambda^\mu$ given by  \eqref{lambda before limit poin bms}, we find the following curve:
 \begin{eqnarray}\label{equation for lambda curve before limit}
	\nonumber	& \ell^2 l_\phi \phi- u l_u + \dfrac{\ell^2 l_u}{r} = 0,\\
	&4 l_u \ell^2 \phi^2 -  8 \ell^2 l_\phi \phi u + \ell^2 l_\phi^2 l_u +4 l_u u^2 - l_u^3 = 0. 			
\end{eqnarray}
 It is clear that \eqref{equation for xi} and \eqref{equation for lambda curve before limit} are changed to each other by using \eqref{transformation of l}. Furthermore, the new curve that we show it by $\gamma_{AdS}$ in this paper, intersect   the boundary at the points $\left(u=-\dfrac{l_\phi\ell}{2}, \phi=-\dfrac{l_u}{2\ell}\right )$ and $\left(u=\dfrac{l_\phi\ell}{2}, \phi=\dfrac{l_u}{2\ell}\right )$. These points can be assumed as the end points of a new timelike interval on the boundary, which is given by transformation \eqref{transformation of l} from the original spacelike interval.  The curve $\gamma_{AdS}$ is the same extremal curve, which is given by the RT proposal for the new interval on the boundary. The interesting point is that both $\gamma_{AdS}$ and RT extremal curves have the same length. This means that their length is invariant under the transformation \eqref{transformation of l}. 
 
 Now, let us look at \eqref{equation for lambda curve before limit} and keep terms which have $\ell^2$. This yields  the following equations:
 \begin{eqnarray}\label{gammaequation for pinc flat}
	 \nonumber &l_\phi \phi+\dfrac{l_u}{r}=0,\\
	 & 4l_u\phi^2-8l_\phi\phi u+l_\phi^2 l_u=~0
 \end{eqnarray}
 These are exactly the same equations, which result from $\xi^\mu_{flat}=0$, where $\xi^\mu_{flat}$ is given by \eqref{modular flow flat poincare}.\footnote{  Only two equations of $\xi^\mu_{flat}=0$ are independent. } One can use \eqref{gammaequation for pinc flat} and write $\phi$ and $u$ in terms of $r$ as
 \begin{eqnarray}\label{gamma in flat}
 	\nonumber &\phi=-\dfrac{l_u}{l_\phi r},\\
 	& u=-\dfrac{l_\phi^2}{8}r-\dfrac{l_u^2}{2 l_\phi^2 r}.
\end{eqnarray}   
These are the equations that describe curve $\gamma$ in \cite{Jiang:2017ecm}. However, $\gamma$ is not the entire curve and occupies only a portion of it, which is determined by the intersection of $\gamma^+$ and $\gamma^-$. Our next task is to determine new curves $\gamma^\pm_{AdS}$ in \eqref{poincare-bms} which their flat-space limit yields $\gamma^{\pm}$. 
 
 To do so, we use the fact that $\gamma^\pm$ are null curves that connect end points of the interval at null infinity to $\gamma$. Thus, we propose $\gamma^\pm_{AdS}$ as those null geodesics which connect end points of the interval at the boundary to $\gamma_{AdS}$ (Fig. 1).
 
\begin{figure}[h]
   \centering
    \includegraphics[width=0.3\textwidth]{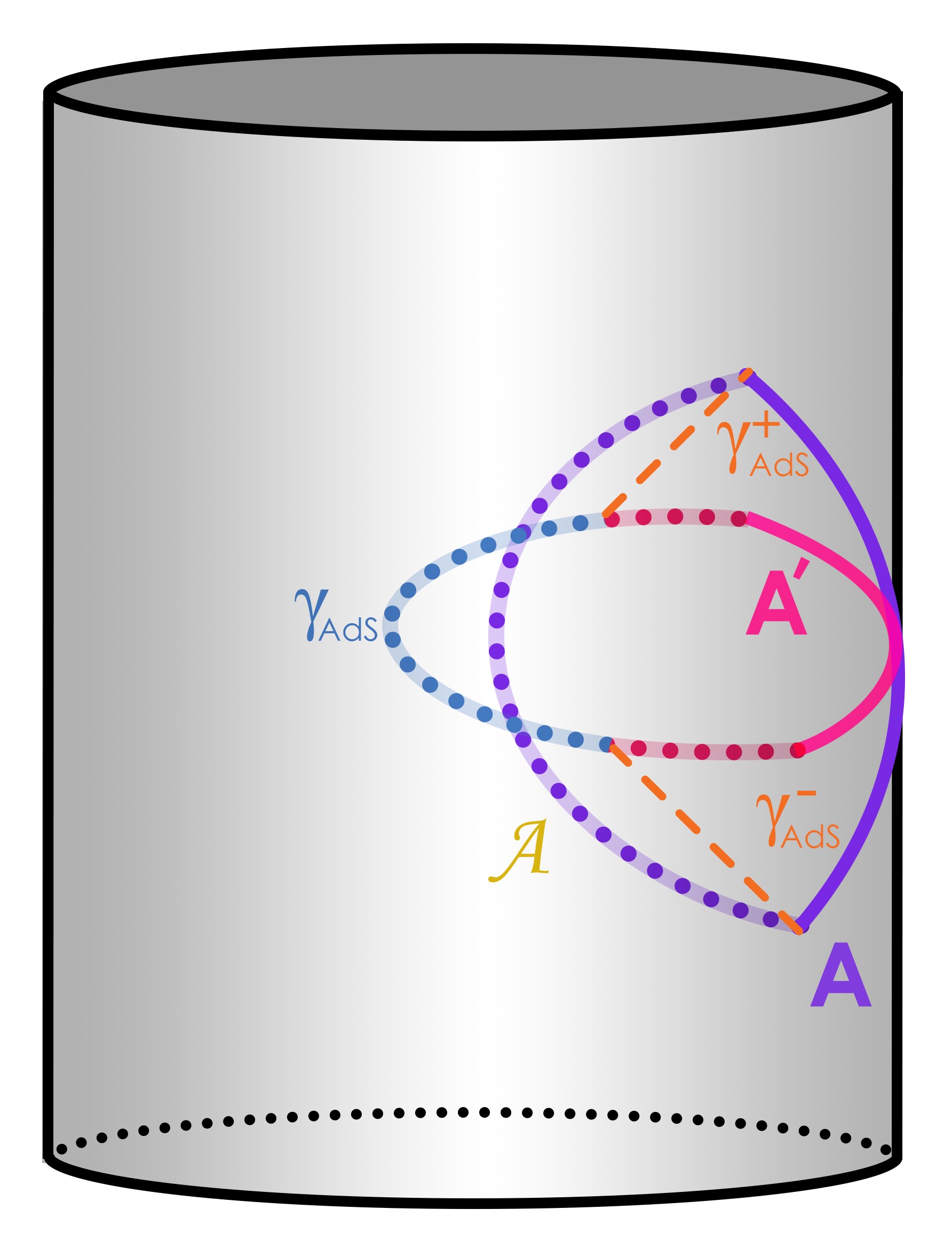} 
 	\caption{Curve   $\mathbf{A} $ is the boundary interval and $\mathcal{A}$ is the extremal curve given by the RT prescription. $\mathbf{A'} $ is the new interval and $\gamma_{AdS}$ is the new extremal curve. $\gamma^\pm_{AdS}$ connects end points of  $\mathbf{A} $ to $\gamma_{AdS}$.}
 	\label{fig1}
\end{figure}

 Starting from geodesic equations of metric \eqref{poincare-bms}, we look for null geodesics that connects $(u=-l_u/2,\phi=-l_\phi/2)$ and $(u=l_u/2,\phi=l_\phi/2)$ at $r=\infty$ to the new curve \eqref{equation for lambda curve before limit}. Solving the geodesic equations with this conditions yields the following results:
 \begin{itemize}
 	\item $\gamma^-_{AdS}$ is a null ray that connects  $(u=-l_u/2,\phi=-l_\phi/2)$ to $\gamma_{AdS}$ and it is determined by equation
 		\begin{equation}\label{equation of gamma m ads}
			 u=-\dfrac{l_u}{2}+\dfrac{\beta_-}{r},\qquad \phi=-\dfrac{l_\phi}{2}+\dfrac{\alpha_-}{r},
 		\end{equation} 
where $\alpha_-$ and $\beta_-$ are two constants determined by using the following equation
		\begin{equation}\label{equation of alpha beta m}
			\alpha_-^2+2\beta_- -\dfrac{\beta_-^2}{\ell^2}=0.
		\end{equation}
 
	 \item $\gamma^+_{AdS}$ is a null geodesic which connects  $(u=l_u/2,\phi=l_\phi/2)$ to $\gamma_{AdS}$. It is given by
		 \begin{equation}\label{equation of gamma p ads}
			 u=\dfrac{l_u}{2}+\dfrac{\beta_+}{r},\qquad \phi=\dfrac{l_\phi}{2}+\dfrac{\alpha_+}{r},
 		\end{equation}
where $\alpha_+$ and $\beta_+$ are two constants, satisfy the following equation
		\begin{equation}\label{equation of alpha beta p}
			\alpha_+^2+2\beta_+ -\dfrac{\beta_+^2}{\ell^2}=0.
		\end{equation}
\end{itemize}
Thus, instead of a unique curve, we spot a bunch of curves for $\gamma^\pm_{AdS}$.  The interesting point is that $\lambda^\mu$ is tangent to all of these curves.  To determine unique null rays emanating from the end points of the boundary interval, we demand that these curves intersects some cutoff points. Precisely, we demand that $\gamma^{+}_{AdS}$ passes through $\left(r=\dfrac{1}{\epsilon_+},u=\dfrac{l_u}{2}-\epsilon_u,\phi=\dfrac{l_\phi}{2}-\epsilon_\phi\right)$ and $\gamma^{-}_{AdS}$ crosses    $\left(u=-\dfrac{l_u}{2}+\epsilon_u,\phi=\dfrac{-l_\phi}{2}\epsilon_\phi\right)$ at $r=\dfrac{1}{\epsilon_-}$, where $\epsilon_u$, $\epsilon_\phi$, and $\epsilon_\pm$ are infinitesimal positive constants and $\epsilon_u/\epsilon_\phi$ is finite. With these conditions, we find that $\beta_\pm=\dfrac{\epsilon_u}{\epsilon_\phi}\alpha_\pm$, and there are two sets of solutions for $\alpha_\pm$:
\begin{align}
 			\begin{cases}
				& \alpha_+=\alpha_-=0  \\
				& \,\,\,\,\,\,\,\,\text{or} \\
				& \alpha_+=\alpha_-=-\dfrac{\dfrac{2\epsilon_u}{\epsilon_\phi}}{1-\dfrac{\epsilon_u^2}{\ell^2\epsilon_\phi^2}}.
			\end{cases}
\end{align}
 For the second case, after taking the flat-space limit, one can deduce that $\epsilon_\pm<0$, which is opposed to our earlier assumption. Thus, we need to choose  $\alpha_+=\alpha_-=\beta_+=\beta_-=0$, the flat-space limit of which results in $\gamma_\pm$ of \cite{Jiang:2017ecm}.

\subsection{Our proposal for calculation of holographic BMSFT entanglement entropy }

In the previous subsections, we proposed a holographic method for the calculation of BMSFT entanglement entropy, which does not use the Rindler method and takes a flat-space limit from the calculations of the AdS/CFT correspondence. In this subsection, we summarize our method and provide a generic recipe:
\begin{itemize}
	 \item In the first step we need to find an asymptotically AdS  spacetime that its flat-space limit is well-defined and    results in our asymptotically flat metric. For this purpose, we need to write our metric and the interval at null infinity in the BMS gauge.
	\item For the asymptotically AdS metric and its corresponding interval, we can use RT prescription or its generalizations to gain an extremal curve in bulk, the length which is proportional to the entanglement entropy of the interval. 
	\item The bulk modular flow is a Killing vector that is vanishing on the extremal curve. So, our next step is finding bulk modular flow by using the extremal curve of the asymptotically AdS case.
	\item Using the bulk modular flow, we look for a new Killing vector which is normal to it, and their norms satisfy \eqref{equation of lambda}. This new vector has a well-defined flat-space limit, which is the bulk modular flow of the corresponding asymptotically flat spacetimes.
	\item The points in bulk in which the components of the new vector are vanishing construct a new extremal curve in the asymptotically AdS spacetime. The flat-space limit of this new curve is well defined and yields a curve in the asymptotically flat spacetime that $\gamma$ is a part of it.
	\item In the next step, we look for the null geodesics that start from end points of the interval and intersect our new extremal curve. The flat-space limit of these curves results in $\gamma_\pm$. However, instead of a unique curve, we encounter a bunch of curves with this property. We remove this ambiguity by demanding that the desired null curve must pass through another point in the spacetime, which we call the cutoff point.

	\item Knowing $\gamma$ and $\gamma_\pm$, we can calculate the length of $\gamma$, which is proportional to the entanglement entropy of the interval in BMSFT.

\end{itemize}

\section{Conclusion} 

 We introduced new curves in the three-dimensional asymptotically AdS spacetimes whose flat-space limit yields the extremal curves in the asymptotically flat spacetimes. These curves can be used for the holographic calculation of the dual BMSFT entanglement entropy. In our proposal, a new vector field, which is normal to the bulk modular flow of the corresponding CFT interval, plays an important role. In fact, instead of the BMSFT interval, we considered a similar CFT interval, and using the RT proposal, we found the extremal curve and also the bulk modular flow in the asymptotically AdS spacetime. Then, using the Killing equation and also normality and norm conditions, we found a new vector field. The points in the spacetime in which this new vector field is vanishing construct a curve with a well-defined flat-space limit. We also have proposed two null rays with a well-defined flat-space limit in the asymptotically AdS spacetime, which connects end points of the interval to the new extremal curve. The flat-space limit of these curves together provides a holographic description for the BMSFT entanglement entropy proposed previously in \cite{Jiang:2017ecm}.
 
 We proposed our method as an alternative method for finding the holographic entanglement entropy of BMSFTs, which does not need to use Rindler transformation. However, in the first part of the current paper, we developed our method by arguing that the new RT-like extremal curves are those curves that map to the inner horizon of BTZ black holes. Using this fact, we could find two main formulas of this paper \eqref{equation of lambda}. In this view, the importance of the inner horizon in this work is to help us find \eqref{equation of lambda}. As a result, the connection with the inner horizon is absent in the steps which we propose as a recipe to calculate the holographic entanglement entropy of BMSFTs. Moreover, in the context of the current paper, before taking the flat-space limit, these new extremal curves are not crucial for extracting new information about the holographic entanglement entropy of CFT intervals. Thus, one can escape from the unclear points which arise by extrapolating the known methods in the AdS/CFT correspondence to the inner horizons.

Our method can be generalized to the higher dimensions. One of the future directions is applying it to find the entanglement entropy of three-dimensional BMSFTs which are dual to the four-dimensional asymptotically flat spacetimes. To do so, we need to overcome some circumstances. In general, modular flows can only be identified with geometric flows generated by Killing vectors when the modular Hamiltonian is local. In three dimensions that any arbitrary interval in the boundary CFT is a line segment, the modular Hamiltonian  is local. In higher dimensions, only spherical entangling regions will generate local modular Hamiltonians, so the interpretation of this modular Hamiltonian as generating a geometric flow, which is along the direction of a Killing vector, breaks down. Thus, the RT surface can no longer be considered as a Killing horizon.

 Moreover, this method can be used to relate the first law of entanglement entropy in the boundary theory to the linearized equation of motion of the bulk theory \cite{Lashkari:2013koa,eom}. In the context of flat/BMSFT, this relation has been studied earlier in \cite{Godet:2019wje,Fareghbal:2019czx} (see also more recent papers \cite{Apolo:2020bld,Apolo:2020qjm}). The main advantage of our method is that one can study this problem by taking the flat-space limit from the known results in the AdS/CFT. For example, the extremal curves related to perturbed geometry in the asymptotically flat spacetime can be found by taking the flat-space limit from the extremal curves in the asymptotically AdS spacetime. We plan to study this problem in our future works.

\subsubsection*{Acknowledgements}
We would like to thank Yousef Izadi for his comments on the manuscript. We are also grateful to Daniel Grumiller and Masoud Gharahi Ghahi for useful discussions.  M. H. is grateful for the hospitality of TU Wien where the last part of the current paper was done. We would also like to thank the referee for his/her useful comments.   This work is supported by  Iran National Science Foundation, Project No. 97017212.

\appendix
\section{ Rindler transformation}

We are looking for a Rindler transformation which changes 
\begin{equation}\label{ap poin bms}
	ds^2=-\dfrac{r^2}{\ell^2}du^2-2du dr+r^2 d\phi^2,
\end{equation}
to a BTZ black hole written in the BMS coordinate,
\begin{equation}\label{btzbms}
	ds^2 = \left( - \frac{\tilde r^2}{\ell^2} + \mathcal{\tilde M}\right)d\tilde u^2 -2d\tilde u d\tilde r +2 \mathcal{\tilde N}d\tilde u d\tilde\phi +\tilde r^2 d\tilde\phi^2,
\end{equation}
where $\mathcal{\tilde M}$ and  $\mathcal{\tilde N}$ are constants and are given in terms of the BTZ inner and outer horizons $\tilde r_\pm$ as
\begin{equation}
	\mathcal{\tilde M}=\dfrac{\tilde r_+^2+\tilde r_-^2}{\ell^2},\qquad \mathcal{\tilde N}=\dfrac{\tilde r_+\tilde r_-}{\ell}.
\end{equation}
To use the transformations which were introduced in the literature, we use  \eqref{boost} and \eqref{poin to bms ads} to change \eqref{ap poin bms} to a Poincar\'e coordinate,
\begin{equation}\label{ap Poin}
	ds^2=\dfrac{\ell^2}{z^2}\left(dz^2-dt^2+dx^2\right).
\end{equation}
Then, using the transformation 
\begin{equation}
	\rho=\dfrac{1}{2z^2},\quad U=x-t, \quad V=t+x,
\end{equation}
we change \eqref{ap Poin} to 
\begin{equation}\label{ap PoinUV}
	ds^2=\ell^2\left(\dfrac{d\rho^2}{4\rho^2}+2\rho dU dV\right).
\end{equation}
The corresponding Rindler transformation is given by (7.2) of \cite{Jiang:2017ecm} as 
\begin{align}\label{ctUuads}
	 T_{\tilde{U}} \tilde{U} =&\frac{1}{4}\log\Big[\frac{(1+ \rho(2 U+l_U) V)^2-  \rho^2 l_V^2(l_U/2+U)^2}{(1+ \rho(2U-l_U) V)^2-  \rho^2 l_V^2(l_U/2-U)^2}\Big]\,,
	\cr
	 T_{\tilde{V}} \tilde{V} =&\frac{1}{4}\log\Big[\frac{(1+ \rho(2 V+l_V) U)^2-  \rho^2 l_U^2(l_V/2+V)^2}{(1+ \rho(2V-l_V) U)^2-  \rho^2 l_U^2(l_V/2-V)^2}\Big]\,,
	\cr
 	\frac{\tilde{\rho}}{T_{\tilde{U}} T_{\tilde{V}}} =&\frac{\rho ^2 \left(l_U^2 \left(l_V^2-4 V^2\right)-4 U^2 l_V^2\right)+4 (2 \rho  U V+1)^2}{4 \rho  l_U l_V}\,,
\end{align}
and transforms \eqref{ap PoinUV} to 
\begin{align}\label{btztutv}
	ds^2&=\ell^2\left(T_{\tilde{U}}^2 \,d \tilde{U} ^2+2  \tilde{\rho}   \,d \tilde{U}  d \tilde{V} + T_{\tilde{V}}^2 d \tilde{V} ^2 + \frac{d \tilde{\rho}  ^2 }{4 ( \tilde{\rho} ^2 -T_{\tilde{U}}^2 T_{\tilde{V}}^2)}\right)\,.
\end{align}
The final part of the transformation is given by 
\begin{align}\label{dipoleBMS}
	\tilde U=& \frac{ \tanh ^{-1}\left(\frac{2 \tilde r}{T_{\tilde V} -T_{\tilde U} }\right)-\tanh ^{-1}\left(\frac{2 \tilde r}{T_{\tilde V} +T_{\tilde U} }\right)}{2  T_{\tilde U} }+\frac{\ell\tilde\phi+\tilde u}{2\ell^2}\,,
	\cr
	\tilde V=& \frac{ \tanh ^{-1}\left(\frac{2 \tilde r}{T_{\tilde V} -T_{\tilde U} }\right)+\tanh ^{-1}\left(\frac{2 \tilde r}{T_{\tilde V} +T_{\tilde U} }\right)}{2  T_{\tilde V} }+\frac{\ell\tilde\phi-\tilde u}{2\ell^2}\,,
	\cr
	\tilde \rho =& \frac{1}{2} \left(4\tilde r^2-T_{\tilde V} ^2-T_{\tilde U} ^2\right)\,,
\end{align}
which transforms \eqref{btztutv} to \eqref{btzbms}, and we have
\begin{equation}
	T_{\tilde U}=\tilde r_+ +\tilde r_-,\qquad T_{\tilde V}=\tilde r_+ -\tilde r_-. 
\end{equation}

\section{Holographic entanglement entropy of BMSFT dual to global Minkowski}

Let us consider an interval in BMSFT$_2$ dual to the three-dimensional global Minkowski spacetime given by the following metric:
\begin{equation}\label{global Minkowski}
 	ds^2=-d\hat u^2-2d\hat u d\hat r+\hat r^2 d\hat\phi^2.
\end{equation} 
The interval is determined by 
\begin{equation}\label{interval}
	\hat u=\dfrac{L_u\sin\hat\phi}{2\sin \dfrac{L_\phi}{2}},
\end{equation}
and end points of it are at $(\hat u=-L_u/2, \hat\phi=-L_\phi/2)$ and $(\hat u=+L_u/2, \hat\phi=+L_\phi/2)$. 

The metric \eqref{global Minkowski} is given by taking the flat-space limit from the global AdS written in the BMS gauge,
\begin{equation}\label{global AdS}
	 ds^2=-\left(1+\dfrac{\hat r^2}{\ell^2}\right)d\hat u^2-2d\hat u d\hat r+\hat r^2 d\hat\phi^2.
\end{equation}
To find the holographic entanglement of the interval \eqref{interval} in BMSFT, we start with the same interval in the dual CFT$_2$ of \eqref{global AdS}. The bulk modular flow, $\xi^\mu$, and the new vector $\lambda^\mu$ are given by using  the following transformation from \eqref{bulk modular flow BMS ads} and \eqref{lambda before limit poin bms}:

\begin{eqnarray}\label{poin to gobal ads}
	\nonumber & r=\frac{1}{2}\left(\ell\sin\dfrac{\hat u}{\ell}+\hat r\cos\dfrac{\hat u}{\ell}+\hat r\cos \hat\phi\right),\\
	\nonumber & \phi=\dfrac{\hat r \sin \hat\phi}{r},\\
	& u=\dfrac{\ell}{r} \left(\ell-\ell\cos\dfrac{\hat u}{\ell}+\hat r \sin\dfrac{\hat u}{\ell}\right).
\end{eqnarray} 
The transformation \eqref{poin to gobal ads} determines $l_\phi$ and $l_u$ in terms of $L_\phi$ and $L_u$,
\begin{eqnarray}
	 l_u=\frac{4 \ell  \sin \left(\frac{L_u}{2 \ell }\right)}{\cos \left(\frac{L_u}{2 \ell }\right)+\cos \left(\frac{L_{\phi }}{2}\right)},\qquad  l_\phi=\frac{4\sin \left(\frac{L_{\phi }}{2}\right)}{\cos \left(\frac{L_u}{2 \ell }\right)+\cos \left(\frac{L_{\phi }}{2}\right)},
\end{eqnarray}
and gives $\xi^\mu$ and $\lambda ^\mu$ of the global AdS as follows:
\begin{eqnarray}\label{xi lambda global}
	\nonumber 
	& \xi=\frac{4 \pi  \left(\cos \left(\hat{\phi }\right) \sin \left(\frac{L_{\phi }}{2}\right) \cos \left(\frac{L_u}{2 \ell }\right) \left(\ell  \cos \left(\frac{\hat{u}}{\ell }\right)-\hat{r} \sin \left(\frac{\hat{u}}{\ell }\right)\right)+\sin \left(\hat{\phi }\right) \cos \left(\frac{L_{\phi }}{2}\right) \sin \left(\frac{L_u}{2 \ell }\right) \left(\hat{r} \cos \left(\frac{\hat{u}}{\ell }\right)+\ell  \sin \left(\frac{\hat{u}}{\ell }\right)\right)\right)}{\cos \left(L_{\phi }\right)-\cos \left(\frac{L_u}{\ell }\right)} \partial_{\hat{r}} +\\
	& \frac{2 \pi  \ell  \left(-2 \cos \left(\hat{\phi }\right) \sin \left(\frac{L_{\phi }}{2}\right) \cos \left(\frac{\hat{u}}{\ell }\right) \cos \left(\frac{L_u}{2 \ell }\right)-2 \sin \left(\hat{\phi }\right) \cos \left(\frac{L_{\phi }}{2}\right) \sin \left(\frac{\hat{u}}{\ell }\right) \sin \left(\frac{L_u}{2 \ell }\right)+\sin \left(L_{\phi }\right)\right)}{\cos \left(L_{\phi }\right)-\cos \left(\frac{L_u}{\ell }\right)} \partial_{\hat{u}} + \\ \nonumber
& \frac{2 \pi  \left(2 \cos \left(\hat{\phi }\right) \cos \left(\frac{L_{\phi }}{2}\right) \sin \left(\frac{L_u}{2 \ell }\right) \left(\hat{r} \cos \left(\frac{\hat{u}}{\ell }\right)+\ell  \sin \left(\frac{\hat{u}}{\ell }\right)\right)+2 \sin \left(\hat{\phi }\right) \sin \left(\frac{L_{\phi }}{2}\right) \cos \left(\frac{L_u}{2 \ell }\right) \left(\hat{r} \sin \left(\frac{\hat{u}}{\ell }\right)-\ell  \cos \left(\frac{\hat{u}}{\ell }\right)\right)-\hat{r} \sin \left(\frac{L_u}{\ell }\right)\right)}{\hat{r} \left(\cos \left(L_{\phi }\right)-\cos \left(\frac{L_u}{\ell }\right)\right)}\partial_{\hat{\phi}}, \\ \nonumber
	& \lambda= -\frac{4 \pi  \left(\cos \left(\hat{\phi }\right) \cos \left(\frac{L_{\phi }}{2}\right) \sin \left(\frac{L_u}{2 \ell }\right) \left(\ell  \cos \left(\frac{\hat{u}}{\ell }\right)-\hat{r} \sin \left(\frac{\hat{u}}{\ell }\right)\right)+\sin \left(\hat{\phi }\right) \sin \left(\frac{L_{\phi }}{2}\right) \cos \left(\frac{L_u}{2 \ell }\right) \left(\hat{r} \cos \left(\frac{\hat{u}}{\ell }\right)+\ell  \sin \left(\frac{\hat{u}}{\ell }\right)\right)\right)}{\cos \left(L_{\phi }\right)-\cos \left(\frac{L_u}{\ell }\right)} \partial_{\hat{r}} + \\
	& \frac{2 \pi  \ell  \left(2 \cos \left(\hat{\phi }\right) \cos \left(\frac{L_{\phi }}{2}\right) \cos \left(\frac{\hat{u}}{\ell }\right) \sin \left(\frac{L_u}{2 \ell }\right)+2 \sin \left(\hat{\phi }\right) \sin \left(\frac{L_{\phi }}{2}\right) \sin \left(\frac{\hat{u}}{\ell }\right) \cos \left(\frac{L_u}{2 \ell }\right)-\sin \left(\frac{L_u}{\ell }\right)\right)}{\cos \left(L_{\phi }\right)-\cos \left(\frac{L_u}{\ell }\right)} \partial_{\hat{u}} +\\ \nonumber
	& \frac{2 \pi  \left(-2 \cos \left(\hat{\phi }\right) \sin \left(\frac{L_{\phi }}{2}\right) \cos \left(\frac{L_u}{2 \ell }\right) \left(\hat{r} \cos \left(\frac{\hat{u}}{\ell }\right)+\ell  \sin \left(\frac{\hat{u}}{\ell }\right)\right)+2 \sin \left(\hat{\phi }\right) \cos \left(\frac{L_{\phi }}{2}\right) \sin \left(\frac{L_u}{2 \ell }\right) \left(\ell  \cos \left(\frac{\hat{u}}{\ell }\right)-\hat{r} \sin \left(\frac{\hat{u}}{\ell }\right)\right)+\hat{r} \sin \left(L_{\phi }\right)\right)}{\hat{r} \left(\cos \left(L_{\phi }\right)-\cos \left(\frac{L_u}{\ell }\right)\right)}\partial_{\hat{\phi}}.  
\end{eqnarray}
It is clear that these two vector fields transform to each other by using the transformation
\begin{align}\label{transformation of L global}
	  L_u\Rightarrow \ell\,L_\phi,\qquad L_\phi&\Rightarrow\dfrac{L_u}{\ell}.
\end{align}
The rest of the calculation is straightforward, and one can find $\gamma_{AdS}$  and also $\gamma_\pm$ and take the flat-space limit from them to obtain $\gamma$ and $\gamma_\pm$ in the global Minkowski spacetime. The final results are the same as in \cite{Jiang:2017ecm}.

%


\end{document}